\documentclass[prd,twocolumn,groupedaddress,showpacs,nofootinbib]{revtex4}
\usepackage{graphicx}
\usepackage{dcolumn}
\usepackage{amssymb}
\usepackage{mathrsfs}
\usepackage{amsmath}
\usepackage{epsfig}
\usepackage[dvips]{color}
\usepackage{hhline}

\begin{document}

\title{ Spontaneous mirror left-right symmetry breaking for leptogenesis parametrized by Majorana neutrino mass matrix }

\author{Pei-Hong Gu}
\email{peihong.gu@sjtu.edu.cn}

\affiliation{Department of Physics and Astronomy, Shanghai Jiao Tong
University, 800 Dongchuan Road, Shanghai 200240, China}

\begin{abstract}

We introduce a mirror copy of the ordinary fermions and Higgs scalars for embedding the $SU(2)_L^{}\times U(1)_Y^{}$ electroweak gauge symmetry into an $SU(2)_L^{}\times SU(2)_R^{}\times U(1)_{B-L}^{}$ left-right gauge symmetry. We then show the spontaneous left-right symmetry breaking can automatically break the parity symmetry motivated by solving the strong CP problem. Through the $SU(2)_R^{}$ gauge interactions, a mirror Majorana neutrino can decay into a mirror charged lepton and two mirror quarks. Consequently we can obtain a lepton asymmetry stored in the mirror charged leptons. The Yukawa couplings of the mirror and ordinary charged fermions to a dark matter scalar then can transfer the mirror lepton asymmetry to an ordinary lepton asymmetry which provides a solution to the cosmic baryon asymmetry in association with the $SU(2)^{}_L$ sphaleron processes. In this scenario, the baryon asymmetry can be well described by the neutrino mass matrix up to an overall factor.

\end{abstract}


\maketitle

\section{Introduction}

The precise measurements on atmospheric, solar, accelerator and reactor neutrinos have established the phenomena of neutrino oscillations. This discovery implies three flavors of neutrinos should be massive and mixed \cite{patrignani2016}. However, the neutrinos are massless in the $SU(3)_c^{}\times SU(2)_L^{}\times U(1)^{}_{Y}$ standard model (SM). Furthermore, in the SM context we cannot explain the cosmic matter-antimatter asymmetry which is as same as a baryon asymmetry \cite{patrignani2016}. Currently a seesaw \cite{minkowski1977,mw1980,flhj1989} extension of the SM has become very attractive since it can simultaneously produce the small neutrino masses and the cosmic baryon asymmetry \cite{fy1986,lpy1986,fps1995,ms1998,bcst1999,hambye2001,di2002,gnrrs2003,hs2004,bbp2005,adjlr2006,dnn2008}. The usual seesaw models contain some heavy ingredients with lepton-number-violating interactions. The neutrinos then can obtain a Majorana mass term which is highly suppressed by a small ratio of the electroweak scale over the heavy masses of the new fields. Meanwhile, these heavy fields can decay to generate a lepton asymmetry through their CP-violating Yukawa and/or scalar interactions. Subsequently, this lepton asymmetry can be partially converted to a baryon asymmetry through the sphaleron processes \cite{krs1985}. This baryogensis scenario in the seesaw models is the so-called leptogenesis mechanism. In such leptogenesis scenario, we do not know much about the texture of the masses and couplings involving the newly heavy fields. Consequently, we cannot get an exact relation between the cosmic baryon asymmetry and the neutrino mass matrix. For example, we can expect a successful leptogenesis in some seesaw models even if the neutrino mass matrix does not contain any CP phases \cite{di2001,gu2016}.

The seesaw models can naturally originate from more fundamental theories beyond the SM, such as the $SU(3)_c^{}\times SU(2)_L^{}\times SU(2)_R^{}\times U(1)_{B-L}^{}$ \cite{ps1974} left-right symmetric theory. In the most popular left-right symmetric models, the SM left-handed fermions are placed in the $SU(2)_L^{}$ doublets as they are in the SM while the SM right-handed fermions plus three right-handed neutrinos are placed in the $SU(2)_R^{}$ doublets. After the left-right symmetry is spontaneously broken down to the electroweak symmetry, we can obtain the lepton number violation for the seesaw and leptogenesis
mechanisms. For example, the right-handed neutrinos can acquire a heavy Majorana mass term to violate the lepton number by two units. Alternatively, the left-right symmetric framework can offer a universal seesaw scenario \cite{berezhiani1983,bm1989,bcs1991,gm2017}, where some additional $[SU(2)]$-singlet fermions with heavy masses are introduced to construct the Yukawa couplings with the $[SU(2)]$-doublet fermions and Higgs. By integrating out these heavy fermion singlets, the SM fermions including the left-handed neutrinos can obtain the desired masses. In the universal seesaw models, the strong CP problem even can be solved without an unobserved axion if a discrete parity symmetry is imposed \cite{bm1989,bcs1991,ms1978,lavoura1997}. 

The $SU(3)_c^{}\times SU(2)_L^{}\times SU(2)_R^{}\times U(1)_{B-L}^{}$ left-right symmetry can be realized in another way \cite{bcs1991,gu2012,cgnr2013,gu2014,abbas2017} where the SM left-handed fermions are the $SU(2)^{}_L$ doublets, the SM right-handed fermions are the $SU(2)$ singlets, meanwhile, the ordinary SM fermions have a mirror copy gauged by the $SU(3)_c^{}\times SU(2)^{}_{R}\times U(1)^{}_{B-L}$ group. The ordinary and mirror fermions can have no mixing because of certain unbroken discrete symmetry \cite{gu2012,gu2014}. Instead, the mirror fermions can decay into the ordinary fermions with a dark matter scalar \cite{gu2012,gu2014}. In such type of mirror left-right symmetric models, the mirror Majorana or Dirac neutrino decays induced by the $SU(2)^{}_R$ gauge interactions can generate a lepton asymmetry stored in some flavors of mirror charged leptons. All or part of this mirror lepton asymmetry then can be transferred to an ordinary lepton asymmetry for a successful leptogenesis. In the early works \cite{gu2012,gu2014}, a softly broken parity symmetry is imposed so that the Yukawa couplings for the mirror fermion mass generation can be identified with those for the ordinary fermion mass generation. In consequence, the cosmic baryon asymmetry can have a distinct dependence on the neutrino mass matrix.

In this paper, we shall show in the mirror left-right symmetric framework, the spontaneous left-right symmetry breaking can automatically break the parity symmetry motivated by solving the strong CP problem. Under this parity, not only the dimensionless Yukawa couplings but also the Majorana masses of the gauge-singlet fermions are identified in the ordinary and mirror sectors. The gauge-singlet fermions can accommodate the type-I seesaw for generating the Majorana masses of the ordinary left-handed neutrinos and the mirror right-handed neutrinos. Every type of mirror fermions thus can have a mass matrix proportional to that of their ordinary partners. Through the mirror Majorana neutrino decays mediated by the $SU(2)_R^{}$ gauge interactions, we can parametrize the cosmic baryon asymmetry by the neutrino mass matrix up to an overall factor.

\section{The model}

Besides the $SU(3)_c^{}$, $SU(2)_{L}^{}$, $SU(2)_{R}^{}$ and $U(1)_{B-L}^{}$ gauge fields, i.e. $G^{1,...,8}_\mu$, $W^{1,2,3}_{L\mu}$, $W^{1,2,3}_{R\mu}$ and $B^{}_{\mu}$, the model contains the fermions and scalars as below,
\begin{eqnarray}
\!\!\!\!\!\!\!\!\!\!\!\!&&\begin{array}{rcl}
q^{}_L(3,2,1,+\frac{1}{3})(-1)(-,+)\!\!\!\!&\stackrel{P}{\longleftrightarrow}& \!\!q'^{}_R(3,1,2,+\frac{1}{3})(-1)(+,-)\,,\\
[2mm]
d^{}_R(3,1,1,-\frac{2}{3})(+1)(-,+)\!\!\!\!&\stackrel{P}{\longleftrightarrow} &\!\!d'^{}_L(3,1,1,-\frac{2}{3})(+1)(+,-)\,,\\
[2mm]
u^{}_R(3,1,1,+\frac{4}{3})(+1)(-,+)\!\!\!\!&\stackrel{P}{\longleftrightarrow} &\!\!u'^{}_L(3,1,1,+\frac{4}{3})(+1)(+,-)\,,\\
[2mm]
l^{}_L(1,2,1,-1)(-1)(-,+)\!\!\!\!&\stackrel{P}{\longleftrightarrow} &\!\!l'^{}_R(1,1,2,-)(-1)(+,-)\,,\\
[2mm]
e^{}_R(1,1,1,-2)(+1)(-,+)\!\!\!\!&\stackrel{P}{\longleftrightarrow}&\!\! e'^{}_L(1,1,1,-2)(+1)(+,-)\,,\\
[2mm]
N^{}_R(1,1,1,0)(+1)(-,+)\!\!\!\!&\stackrel{P}{\longleftrightarrow}&\!\! N'^{}_L(1,1,1,0)(+1)(+,-)\,,\\
[2mm]
\phi^{}_1(1,2,1,+1)(-2)(+,+)\!\!\!\!&\stackrel{P}{\longleftrightarrow}&\!\! \phi'^{}_1(1,1,2,+1)(-2)(+,+)\,,\\
[2mm]
\phi^{}_2(1,2,1,+1)(+2)(+,+)\!\!\!\!&\stackrel{P}{\longleftrightarrow} &\!\!\phi'^{}_2(1,1,2,+1)(+2)(+,+)\,,\\
[2mm]
\chi(1,1,1,0)(0)(-,-)\!\!\!\!&\stackrel{P}{\longleftrightarrow} &\!\!\chi(1,1,1,0)(0)(-,-)\,.
\end{array}\nonumber\\
\!\!\!\!\!\!\!\!\!\!\!\!&&
\end{eqnarray}
Here the first, second and third brackets following the fields describe the transformations under the $SU(3)^{}_c\times SU(2)^{}_L\times SU(2)^{}_R \times U(1)^{}_{B-L}$ gauge groups, a $U(1)_G^{}$ global symmetry and a $Z^{}_2\times Z^{}_2$ discrete symmetry, respectively. The $SU(2)^{}_L\times SU(2)^{}_R \times U(1)^{}_{B-L}$ left-right symmetry will be spontaneously broken down to the $SU(2)^{}_L \times U(1)^{}_{Y}$ electroweak symmetry and then the $U(1)_{em}^{}$ electromagnetic symmetry, the $U(1)^{}_G$ global symmetry will be assumed softly broken, while the $Z^{}_2\times Z^{}_2$ symmetry will be required conserved at any scales.

The full Lagrangian includes the kinetic terms, the scalar potential and the Yukawa couplings as usual. Furthermore, the gauge-singlet fermions $N^{}_{R}$ and $N'^{}_{L}$ are allowed to have the Majorana masses. For simplicity, we will not show the kinetic terms where the $SU(2)^{}_{R}$ gauge coupling $g^{}_R$ is identified with the $SU(2)^{}_L$ gauge coupling $g^{}_L$, i.e. $g^{}_R = g^{}_L= g$ due to the parity symmetry. We now write down the scalar potential, the Yukawa couplings and the Majorana masses. Specifically, the scalar potential is  
\begin{eqnarray}
V&=&\mu^{2}_{1}\left(\phi^{\dagger}_{1}\phi^{}_{1}+\phi'^{\dagger}_{1}\phi'^{}_{1}\right)+\mu^{2}_{2}\left(\phi^{\dagger}_{2}\phi^{}_{2}+\phi'^{\dagger}_{2}\phi'^{}_{2}\right)\nonumber\\
[2mm]
&&+\mu^{2}_{12}\left[\left(\phi^{\dagger}_{1}\phi^{}_{2}+\phi'^{\dagger}_{1}\phi'^{}_{2}\right)+\textrm{H.c.}\right]+\lambda^{}_{1}\left[\left(\phi^{\dagger}_{1}\phi^{}_{1}\right)^2_{}\right.\nonumber\\
[2mm]
&&\left.+\left(\phi'^{\dagger}_{1}\phi'^{}_{1}\right)^2_{}\right]+\lambda^{}_{2}\left[\left(\phi^{\dagger}_{2}\phi^{}_{2}\right)^2_{}+\left(\phi'^{\dagger}_{2}\phi'^{}_{2}\right)^2_{}\right]\nonumber\\
[2mm]
&&+2\lambda^{}_{3}\left(\phi^{\dagger}_{1}\phi^{}_{1}\phi^{\dagger}_{2}\phi^{}_{2}+\phi'^{\dagger}_{1}\phi'^{}_{1}\phi'^{\dagger}_{2}\phi'^{}_{2}\right)+2\lambda^{}_{4}\left(\phi^{\dagger}_{1}\phi^{}_{2}\phi^{\dagger}_{2}\phi^{}_{1}\right.\nonumber\\
[2mm]
&&\left.+\phi'^{\dagger}_{1}\phi'^{}_{2}\phi'^{\dagger}_{2}\phi'^{}_{1}\right)+2\kappa^{}_{1}\phi^{\dagger}_{1}\phi^{}_{1}\phi'^{\dagger}_{1}\phi'^{}_{1}+2\kappa^{}_{2}\phi^{\dagger}_{2}\phi^{}_{2}\phi'^{\dagger}_{2}\phi'^{}_{2}\nonumber\\
[2mm]
&&+2\kappa^{}_{3}\left(\phi^{\dagger}_{1}\phi^{}_{1}\phi'^{\dagger}_{2}\phi'^{}_{2}+\phi'^{\dagger}_{1}\phi'^{}_{1}\phi^{\dagger}_{2}\phi^{}_{2}\right)+\frac{1}{2}\mu^{2}_{\chi}\chi^2_{}\nonumber\\
[2mm]
&&+\frac{1}{4}\lambda^{}_{\chi}\chi^4_{}+\xi^{}_1\left(\phi^{\dagger}_{1}\phi^{}_{1}+\phi'^{\dagger}_{1}\phi'^{}_{1}\right)\chi^2_{}+\xi^{}_2\left(\phi^{\dagger}_{2}\phi^{}_{2}\right.\nonumber\\
[2mm]
&&\left.+\phi'^{\dagger}_{2}\phi'^{}_{2}\right)\chi^2_{}\,,
\end{eqnarray}
where the parity symmetry and the $Z^{}_2\times Z^{}_2$ symmetry are exactly conserved while the $U(1)^{}_{G}$ global symmetry is softly broken by the $\mu^{2}_{12}$-term. The Yukawa couplings and Majorana masses also respect the parity symmetry and the $Z^{}_2\times Z^{}_2$ symmetry, i.e. 
\begin{eqnarray}
\mathcal{L}_{Y+M}^{}&=&  -\bar{y}_d^{}\left(\bar{q}^{}_L \phi^{}_1 d^{}_R + \bar{q}'^{}_R \phi'^{}_1 d'^{}_L \right)-f^{}_d\chi \bar{d}^{}_R d'^{}_L\nonumber\\
[2mm]
&&-\bar{y}_u^{}\left(\bar{q}^{}_L \tilde{\phi}^{}_2 u^{}_R + \bar{q}'^{}_R \tilde{\phi}'^{}_2 u'^{}_L \right)-f^{}_u\chi \bar{u}^{}_R u'^{}_L\nonumber\\
[2mm]
&&-\bar{y}_e^{}\left(\bar{l}^{}_L \phi^{}_1 e^{}_R + \bar{l}'^{}_R \phi'^{}_1 e'^{}_L \right)-f^{}_e\chi \bar{e}^{}_R e'^{}_L\nonumber\\
[2mm]
&&-\bar{y}_\nu^{}\left(\bar{l}^{}_L \tilde{\phi}^{}_2 N^{}_R + \bar{l}'^{}_R \tilde{\phi}'^{}_2 N'^{}_L \right)-f^{}_N\chi \bar{N}^{}_R N'^{}_L\nonumber\\
[2mm]
&&-\frac{1}{2}M^{}_N\left(\bar{N}^{c}_R N^{}_R +\bar{N}'^{c}_L N'^{}_L \right)+\textrm{H.c.}\,.
\end{eqnarray}
Here the Majorana masses $M^{}_N=M^{T}_N$ also break the  $U(1)^{}_{G}$ global symmetry.

\section{Spontaneous symmetry breaking}

We expect the $SU(2)^{}_L\times SU(2)^{}_R \times U(1)^{}_{B-L}$ left-right symmetry will be spontaneously broken down to the $SU(2)^{}_L \times U(1)^{}_{Y}$ electroweak symmetry. Subsequently, the $SU(2)^{}_L \times U(1)^{}_{Y}$ electroweak symmetry will be spontaneously broken down to the $U(1)^{}_{em}$ electromagnetic symmetry. For demonstration, we denote the VEVs by
\begin{eqnarray}
\langle\phi^{}_{1,2}\rangle &=&\frac{1}{\sqrt{2}}v^{}_{1,2}\,,~~\,v=\sqrt{v^{2}_{1}+v^{2}_{2}}\simeq 246\,\textrm{GeV}\,;\nonumber\\
[2mm]
\langle\phi'^{}_{1,2}\rangle &=&\frac{1}{\sqrt{2}}v'^{}_{1,2}\,,~~v'=\sqrt{v'^{2}_{1}+v'^{2}_{2}}\,.
\end{eqnarray}
Furthermore, we assume the $Z^{}_2\times Z^{}_2$ discrete symmetry will not be broken at any scales. This can be achieved by 
\begin{eqnarray}
m_\chi^2 = \mu_\chi^2 + \xi_{1}^{}(v^{2}_{1}+v'^{2}_{1})+ \xi_{2}^{}(v^{2}_{2}+v'^{2}_{2})>0\,,~~\lambda_\chi^{}>0\,.
\end{eqnarray}

We now derive the VEVs $v^{}_{1,2}$ and $v'^{}_{1,2}$. From the extreme conditions, 
\begin{eqnarray}
\frac{\partial V}{\partial v^{}_1} =\frac{\partial V}{\partial v^{}_2} =\frac{\partial V}{\partial v'^{}_1} =\frac{\partial V}{\partial v'^{}_2}=0 \,,
\end{eqnarray}
we can read
\begin{eqnarray}
\!\!\!\!\!\!\!\!&&\left[\mu^{2}_{1}+ \lambda^{}_{1} v^{2}_{1} +( \lambda^{}_{3}+\lambda^{}_{4})v^{2}_{2} +\kappa^{}_{1}v'^{2}_{1}+\kappa^{}_{3}v'^{2}_{2}\right]v^{}_{1}+ \mu^{2}_{12} v^{}_{2}\nonumber\\
\!\!\!\!\!\!\!\!&&=0\,,\nonumber\\
[2mm]
\!\!\!\!\!\!\!\!&&\left[\mu^{2}_{2}+ \lambda^{}_{2} v^{2}_{2} +( \lambda^{}_{3}+\lambda^{}_{4})v^{2}_{1} +\kappa^{}_{2}v'^{2}_{2}+\kappa^{}_{3}v'^{2}_{1}\right]v^{}_{2}+ \mu^{2}_{12} v^{}_{1}\nonumber\\
\!\!\!\!\!\!\!\!&&=0\,,\nonumber\\
[2mm]
\!\!\!\!\!\!\!\!&&\left[\mu^{2}_{1} + \lambda^{}_{1} v'^{2}_{1} +( \lambda^{}_{3}+\lambda^{}_{4})v'^{2}_{2} +\kappa^{}_{1}v^{2}_{1}+\kappa^{}_{3}v^{2}_{2}\right]v'^{}_{1}+ \mu^{2}_{12} v'^{}_{2}\nonumber\\
\!\!\!\!\!\!\!\!&&=0\,,\nonumber\\
[2mm]
\!\!\!\!\!\!\!\!&&\left[\mu^{2}_{2}+ \lambda^{}_{2} v'^{2}_{2} +( \lambda^{}_{3}+\lambda^{}_{4})v'^{2}_{1} +\kappa^{}_{2}v^{2}_{2}+\kappa^{}_{3}v^{2}_{1}\right]v'^{}_{2}+ \mu^{2}_{12} v'^{}_{1}\nonumber\\
\!\!\!\!\!\!\!\!&&=0\,.
\end{eqnarray}
For a proper parameter choice, the VEVs can have a hierarchical spectrum $v'^{}_{1}> v'^{}_{2}> v^{}_{2}> v^{}_{1}$. This means a spontaneous violation of the parity symmetry \footnote{Similarly we can spontaneously break the mirror symmetry in the models for mirror dark matter.}. For example, in the limiting case $v'^{}_{1}\gg v'^{}_{2}\gg v^{}_{2}\gg v^{}_{1}$, we can analytically solve the VEVs by  
\begin{eqnarray}
v'^{2}_{1}\!\!&\simeq&\!\! -\frac{\mu^{2}_{1}}{\lambda^{}_{1}}\quad\quad\quad\quad\quad\quad\quad~\,\textrm{for}~~\mu^{2}_{1}<0\,,~~\lambda^{}_{1}>0\,;\nonumber\\
[2mm]
v'^{}_{2}\!\!&\simeq &\!\! -\frac{\mu^{2}_{12}v'^{}_{1}}{\mu^{2}_{2}+(\lambda^{}_{3}+\lambda^{}_{4})v'^{2}_{1}}~~~\textrm{for}~~\mu^{2}_{2}+(\lambda^{}_{3}+\lambda^{}_{4})v'^{2}_{1}>0\,;\nonumber\\
[2mm]
v^{2}_{2}\!\!&\simeq& \!\!-\frac{\mu^{2}_{2}+\kappa^{}_{2}v'^{2}_{2}+\kappa^{}_{3}v'^{2}_{1}}{\lambda^{}_{2}}~~\textrm{for}~~\mu^{2}_{2}+\kappa^{}_{2}v'^{2}_{2}+\kappa^{}_{3}v'^{2}_{1}<0\,,\nonumber\\
&&\quad\quad\quad\quad\quad\quad\quad\quad\quad\quad\quad\,\lambda^{}_{2}>0\,;\nonumber\\
[2mm]
v^{}_{1}\!\!&\simeq& \!\!- \frac{\mu^{2}_{12}v^{}_{2}}{\mu^{2}_{1}+\kappa^{}_{1}v'^{2}_{1}+\kappa^{}_{3}v'^{2}_{2}}~~\textrm{for}~~\mu^{2}_{1}+\kappa^{}_{1}v'^{2}_{1}+\kappa^{}_{3}v'^{2}_{2}>0\,.\nonumber\\
\!\!&&
\end{eqnarray}

After the above spontaneous left-right symmetry breaking, the charged fermion mass matrices can be easily given by  
\begin{eqnarray}
\mathcal{L}&\supset& -\left[\bar{d}^{}_{L}~~\bar{d}'^{}_{L}\right]\left[\begin{array}{cc}m^{}_{d}&0\\
[2mm]
0& m'^{\dagger}_{d} \end{array}\right]\left[\begin{array}{c} d^{}_{R}\\
 [2mm]
 d'^{}_{R}\end{array}\right]\nonumber\\
 [2mm] 
 &&-\left[\bar{u}^{}_{L}~~\bar{u}'^{}_{L}\right]\left[\begin{array}{cc}m^{}_{u}&0\\
[2mm]
0& m'^{\dagger}_{u} \end{array}\right]\left[\begin{array}{c} u^{}_{R}\\
 [2mm]
 u'^{}_{R}\end{array}\right]\nonumber\\
 [2mm]
 &&-\left[\bar{e}^{}_{L}~~\bar{e}'^{}_{L}\right]\left[\begin{array}{cc}m^{}_{e}&0\\
[2mm]
0& m'^{\dagger}_{e} \end{array}\right]\left[\begin{array}{c} e^{}_{R}\\
 [2mm]
 e'^{}_{R}\end{array}\right]+\textrm{H.c.}~~\textrm{with}\nonumber\\
 [2mm]
&&m^{}_{d}=\frac{1}{\sqrt{2}}\bar{y}^{}_d v^{}_{1} \propto m'^{}_{d}=\frac{1}{\sqrt{2}}\bar{y}^{}_d v'^{}_{1}\,,\nonumber\\
[2mm]
&&m^{}_{u}=\frac{1}{\sqrt{2}}\bar{y}^{}_u v^{}_{2} \propto m'^{}_{u}=\frac{1}{\sqrt{2}}\bar{y}^{}_u v'^{}_{2}\,,\nonumber\\
[2mm]
&&\,m^{}_{e}=\frac{1}{\sqrt{2}}\bar{y}^{}_e v^{}_{1} \propto m'^{}_{e}=\frac{1}{\sqrt{2}}\bar{y}^{}_e v'^{}_{1}\,.
\end{eqnarray}
As for the neutral fermions, their masses are 
\begin{eqnarray}
\mathcal{L}&\supset& -\frac{1}{2}\left[\bar{\nu}^{}_{L}~~\bar{N}^{c}_{R}\right]\left[\begin{array}{cc}0& \frac{1}{\sqrt{2}}\bar{y}^{}_{N}v^{}_{2}\\
[2mm]
 \frac{1}{\sqrt{2}}\bar{y}^{}_{N} v^{}_{2}& M^{}_{N}\end{array}\right]\left[\begin{array}{c} \nu^{c}_{L}\\
 [2mm]
 N^{}_{R}\end{array}\right]\nonumber\\
 [2mm]
 &&-\frac{1}{2}\left[\bar{\nu}'^{}_{R}~~\bar{N}'^{c}_{L}\right]\left[\begin{array}{cc}0& \frac{1}{\sqrt{2}}\bar{y}^{}_{\nu}v'^{}_{2}\\
[2mm]
 \frac{1}{\sqrt{2}}\bar{y}^{}_{\nu} v'^{}_{2}& M^{}_{N}\end{array}\right]\left[\begin{array}{c} \nu'^{c}_{R}\\
 [2mm]
 N'^{}_{L}\end{array}\right]\nonumber\\
 [2mm]
 &&+\textrm{H.c.}\,.
 \end{eqnarray}
When the seesaw condition is satisfied, i.e. $M_N^{}\gg \bar{y}^{}_{\nu} v^{}_{2}/\sqrt{2}$ and $M_N^{}\gg \bar{y}^{}_{\nu} v'^{}_{2}/\sqrt{2}$, we can integrate out the gauge-singlet fermions $N^{}_{R}$ and $N'^{}_{L}$ to obtain the Majorana masses of the ordinary and mirror neutrinos \footnote{Alternatively the ordinary and mirror neutrinos can obtain their Majorana masses through the type-II \cite{mw1980} or type-III \cite{flhj1989} seesaw. In the case with the type-II seesaw, the Majorana mass matrices of the ordinary and mirror neutrinos can have a same structure even if the parity symmetry is softly broken \cite{gu2012}.}, 
\begin{eqnarray}
\!\!\!\!\!\!\!\!\mathcal{L}&\supset& -\frac{1}{2}\bar{\nu}^{}_{L} m^{}_{\nu} \nu^{c}_{L}- \frac{1}{2} \bar{\nu}'^{}_{R} m'^{}_{\nu}\nu'^{c}_{R} +\textrm{H.c.}~~\textrm{with}\nonumber\\
[2mm]
\!\!\!\!\!\!\!\!&& m^{}_{\nu} = - \bar{y}^{}_{\nu} \frac{v^{2}_{2}}{2M^{}_N} \bar{y}^{T}_{\nu} \propto  m'^{}_{\nu} = - \bar{y}^{}_{\nu} \frac{v'^{2}_{2}}{2M^{}_N} \bar{y}^{T}_{\nu}\,.
\end{eqnarray}

Clearly the mass eigenvalues of the ordinary and mirror fermions obey the relations as below, 
\begin{eqnarray}
&&\frac{m^{}_{d'}}{m^{}_{d}}=\frac{m^{}_{s'}}{m^{}_{s}}=\frac{m^{}_{b'}}{m^{}_{b}}=\frac{v'^{}_{1}}{v^{}_{1}}\,,\nonumber\\
[2mm]
&&\frac{m^{}_{u'}}{m^{}_{u}}=\frac{m^{}_{c'}}{m^{}_{c}}=\frac{m^{}_{t'}}{m^{}_{t}}=\frac{v'^{}_{2}}{v^{}_{2}}\,,\nonumber\\
[2mm]
&&\frac{m^{}_{e'}}{m^{}_{e}}=\frac{m^{}_{\mu'}}{m^{}_{\mu}}=\frac{m^{}_{\tau'}}{m^{}_{\tau}}=\frac{v'^{}_{1}}{v^{}_{1}}\,,\nonumber\\
[2mm]
&&\frac{m^{}_{\nu'^{}_{1}}}{m^{}_{\nu^{}_{1}}}=\frac{m^{}_{\nu'^{}_{2}}}{m^{}_{\nu^{}_{2}}}=\frac{m^{}_{\nu'^{}_{3}}}{m^{}_{\nu^{}_{3}}}=\frac{v'^{2}_{2}}{v^{2}_{2}}\,.
\end{eqnarray}
In the following demonstration, the ordinary mass eigenvalues will be quoted as $m^{}_{t}= 174\,\textrm{GeV}$, $m^{}_{b}= 4.18\,\textrm{GeV}$, $m^{}_{c}= 1.27\,\textrm{GeV}$\,, $m^{}_{s}=96\,\textrm{MeV}$\,, $m^{}_{u}=2.2\,\textrm{MeV}$, $m^{}_{d}=4.7\,\textrm{MeV}$, $m^{}_{\tau}=1.78\,\textrm{GeV}$, $m^{}_{\mu}=106\,\textrm{MeV}$, $m^{}_{e}=511\,\textrm{keV}$, $|\Delta m^{2}_{31}|\equiv |m^{2}_{\nu^{}_{3}}-m^{2}_{\nu^{}_{1}}|\simeq 2.5\times 10^{-3}_{}\,\textrm{eV}^2_{}$, $\Delta m^{2}_{21}\equiv m^{2}_{\nu^{}_{2}}-m^{2}_{\nu^{}_{1}}\simeq 7.5\times 10^{-5}_{}\,\textrm{eV}^2_{}$ \cite{patrignani2016}. Meanwhile, in the mass basis, the charged currents are   
\begin{eqnarray}
\mathcal{L}_{CC}^{}&=&\frac{g}{\sqrt{2}}\left(\bar{u}^{}_{L}\gamma^\mu_{}V d^{}_{L} W^{+}_{L\mu}+\bar{u}'^{}_{R}\gamma^\mu_{}V d'^{}_{R} W^{+}_{R\mu}\right)\nonumber\\
[2mm]
&&+\frac{g}{\sqrt{2}}\left(\bar{e}^{}_{L}\gamma^\mu_{}U^{\ast}_{} \nu^{}_{L} W^{-}_{L\mu}+\bar{e}'^{}_{R}\gamma^\mu_{}U^{\ast}_{} \nu'^{}_{R} W^{-}_{R\mu}\right)\nonumber\\
[2mm]
&&+\textrm{H.c.}\,.
\end{eqnarray} 
Here the charged gauge bosons $W^{\pm}_{L,R}=\frac{1}{\sqrt{2}}\left(W^{1}_{L,R}\mp i W^{2}_{L,R}\right)$ have the masses \cite{patrignani2016},
\begin{eqnarray}
\!\!M^{}_{W^{}_{L}}=\frac{1}{2}gv\simeq 80.4\,\textrm{GeV}\,,~~M^{}_{W^{}_{R}}=\frac{1}{2}gv'\simeq 0.327\,v'^{}\,,
\end{eqnarray} 
the CKM matrix $V$ is simplified by $V\simeq 1$, while the PMNS matrix $U$ is parametrized by \cite{patrignani2016}
\begin{widetext}
\begin{eqnarray}\label{pmns}
U=\left[\begin{array}{ccccl}
c_{12}^{}c_{13}^{}&& s_{12}^{}c_{13}^{}&&  s_{13}^{}e^{-i\delta}_{}\\
[4mm] -s_{12}^{}c_{23}^{}-c_{12}^{}s_{23}^{}s_{13}^{}e^{i\delta}_{}
&&~~c_{12}^{}c_{23}^{}-s_{12}^{}s_{23}^{}s_{13}^{}e^{i\delta}_{}
&& s_{23}^{}c_{13}^{}\\
[4mm] ~~s_{12}^{}s_{23}^{}-c_{12}^{}c_{23}^{}s_{13}^{}e^{i\delta}_{}
&& -c_{12}^{}s_{23}^{}-s_{12}^{}c_{23}^{}s_{13}^{}e^{i\delta}_{}
&& c_{23}^{}c_{13}^{}
\end{array}\right]\textrm{diag}\left\{e^{i\alpha^{}_{1}/2}_{}\,,~e^{i\alpha^{}_{2}/2}_{}\,,1\right\}\,,
\end{eqnarray}
\end{widetext}
with the three mixing angles $\sin^2_{}\theta^{}_{12}\simeq 0.3$, $\sin^2_{}\theta^{}_{23}\simeq 0.5$ and $\sin^2_{}\theta^{}_{13}\simeq 0.02$ \cite{patrignani2016}.

\section{Mirror and ordinary lepton asymmetries}

\begin{figure}
\vspace{9.6cm} \epsfig{file=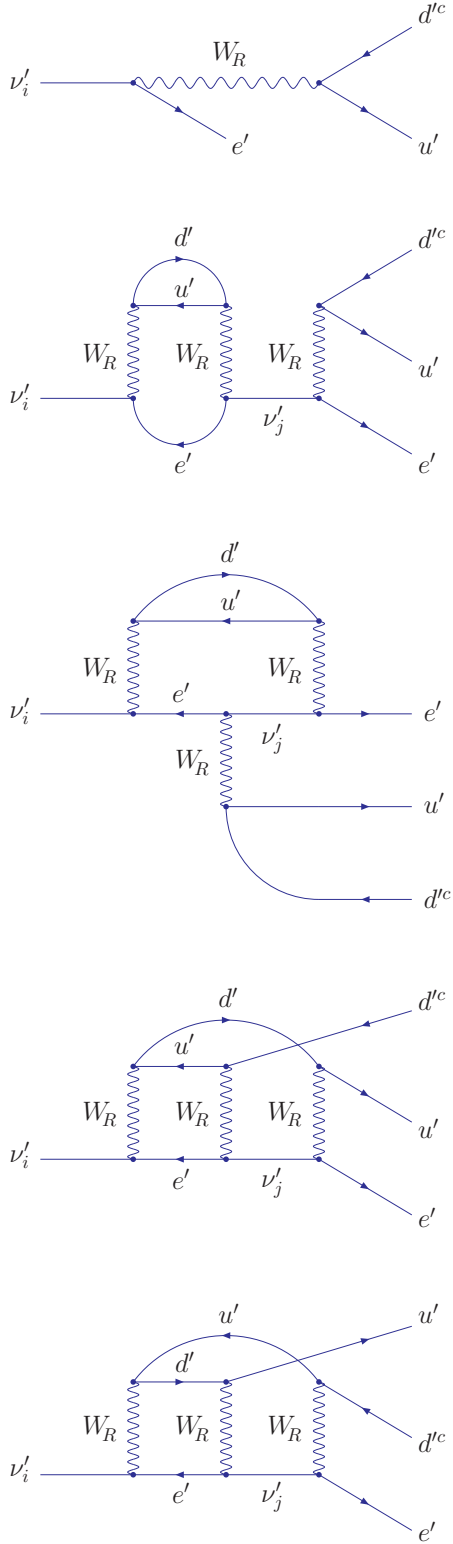, bbllx=-0.5cm, bblly=6.0cm,
bburx=9.5cm, bbury=16cm, width=7cm, height=7cm, angle=0,
clip=0} \vspace{4.5cm} \caption{\label{mndecay} The lepton-number-violating decays of the mirror Majorana neutrinos into the mirror charged fermions. The CP-conjugate processes are not shown for simplicity. }
\end{figure}

When the seesaw-induced Majorana masses of the mirror neutrinos are in the range as below,
\begin{eqnarray}
\label{range}
&&m^{}_{\nu'^{}_i}< M^{}_{W^{}_R}+m^{}_{e'}\,,~m^{}_{\tau'}+m^{}_{u'}+m^{}_{d'}\,,\nonumber\\
&&m^{}_{\nu'^{}_i}\gg m^{}_{\mu'}+m^{}_{u'}+m^{}_{d'}\,, 
\end{eqnarray}
the dominant decay modes of the Majorana mirror neutrinos should be described by Fig. \ref{mndecay}. The mass spectrum (\ref{range}) actually is quite reasonable. By taking the perturbation requirement $\bar{y}_{f}^{}<\sqrt{4\pi}$ into account, the VEV $v^{}_{1}$ should be
\begin{eqnarray}
\frac{m^{}_{b}}{\sqrt{2\pi}}< v^{}_{1}<\sqrt{v^2_{}-\frac{m^{2}_{t}}{2\pi}}\,,
\end{eqnarray}
and hence the Yukawa couplings $\bar{y}_{\tau}^{}$ should be 
\begin{eqnarray}
\frac{m^{}_{\tau}}{m^{}_{t}}\frac{1}{\sqrt{\frac{v^2_{}}{2m^{2}_{t}}-\frac{1}{4\pi}}} < \bar{y}^{}_{\tau}< \frac{m^{}_{\tau}}{m^{}_{b}}\sqrt{4\pi}\,.
\end{eqnarray}
The mirror tau and muon thus should be in the range, 
\begin{eqnarray}
&&0.00694\,v'^{}_{1}< m^{}_{\tau'}<1.07\,v'^{}_{1} \,,\nonumber\\
[2mm]
&&m^{}_{\mu'}=\frac{m^{}_{\mu}}{m^{}_{\tau}}m^{}_{\tau'}=0.0607\,m^{}_{\tau'}\,.
\end{eqnarray}
On the other hand, we can further specify the seesaw condition by $\bar{y}^{}_{\nu}v'^{}_{2}/(\sqrt{2}M^{}_{N})\leq 0.1$ and then put an upper bound on the maximal eigenvalue $m^{}_{\nu'^{}_{3}}$ (normal hierarchy) or $m^{}_{\nu'^{}_{1}}$ (inverted hierarchy) of the mirror neutrino mass matrix $m'^{}_{\nu}$, i.e.
\begin{eqnarray}
m^{}_{\nu'^{}_{3(1)}} < 0.1 \sqrt{2\pi} \,v'^{}_{2}\simeq 0.25\,v'^{}_{2}<M^{}_{W^{}_R}\simeq 0.327\,v'\,. 
\end{eqnarray}
Therefore, we can guarantee the lightest mirror neutrino $\nu'^{}_{1}$ (normal hierarchy) or $\nu'^{}_{3}$ (inverted hierarchy) to math the mass spectrum (\ref{range}) by further taking 
\begin{eqnarray}
\label{v12}
m^{}_{\nu'^{}_{1(3)}}&=&\sqrt{m^{2}_{\nu'^{}_{3(1)}}-\frac{v'^{4}_{2}}{v^{4}_{2}}\left|\Delta m_{31}^2\right|}\nonumber\\
&<&\sqrt{(0.25\,v'^{}_{2})^2_{}-\frac{v'^{4}_{2}}{v^{4}_{2}}\left|\Delta m_{31}^2\right|}< 1.07 \,v'^{}_{1}\,.
\end{eqnarray}
For example, we can input the quasi-degenerate neutrino spectrum in Eq. (\ref{v12}) and then obtain 
\begin{eqnarray}
v'^{}_{1} > \frac{0.25}{1.07}\,v'^{}_{2}\simeq 0.23\,v'^{}_{2}\,,~~v' > 1.02\,v'^{}_{2}\,.
\end{eqnarray}

At tree level, the decay width is calculated by  
\begin{eqnarray}
\Gamma^{}_{\nu'^{}_{i}}&=&\sum_{\alpha k l }^{}\left[ \Gamma (\nu'^{}_{i}\rightarrow e'^{}_{\alpha} + u'^{}_{k} + d'^{c}_{l}) \right.\nonumber\\
[2mm]
&&\left.+ \Gamma (\nu'^{}_{i}\rightarrow e'^{c}_{\alpha} + u'^{c}_{k} + d'^{}_{l}) \right]\nonumber\\
[2mm]
&=& \frac{g^4_{}}{2^{10}\pi^3_{}}\left(\left|U^{}_{ei}\right|^2_{}+\left|U^{}_{\mu i}\right|^2_{}\right)\frac{m^{5}_{\nu'^{}_i}}{M^{4}_{W^{}_R}}\,.
\end{eqnarray}
We can conveniently define 
\begin{eqnarray}
m^{}_{\nu^{}_x}&= &\textrm{min}\{m^{}_{\nu^{}_{1}}\,,~m^{}_{\nu^{}_{2}}\,,~m^{}_{\nu^{}_{3}}\}\,, \nonumber\\
[2mm]
m^{}_{\nu'^{}_x}&=& \textrm{min}\{m^{}_{\nu'^{}_{1}}\,,~m^{}_{\nu'^{}_{2}}\,,~m^{}_{\nu'^{}_{3}}\}\,, ~~r\equiv \frac{m^{2}_{\nu'^{}_x}}{M^{2}_{W^{}_R}}\,,
\end{eqnarray}
and then rewrite the decay width by
\begin{eqnarray}
\Gamma^{}_{\nu'^{}_{i}}=\frac{g^4_{}}{2^{10}\pi^3_{}}\left(\left|U^{}_{ei}\right|^2_{}+\left|U^{}_{\mu i}\right|^2_{}\right)\frac{m^{5}_{\nu^{}_i}}{m^{5}_{\nu^{}_x}} m^{}_{\nu'^{}_x} r^2_{}\,.
\end{eqnarray}
We also compute the CP asymmetry at one-loop level, 
\begin{eqnarray}
\label{cpa}
\varepsilon^{}_{\nu'^{}_{i}}&=&\frac{\left[ \Gamma (\nu'^{}_{i}\rightarrow e' + u'+ d'^{c}_{}) - \Gamma (\nu'^{}_{i}\rightarrow e'^{c}_{} + u'^{c}_{}+ d') \right]}{\Gamma_{i}^{}}\nonumber\\
[2mm]
&=& \frac{g^4_{}}{2^{10}_{}\pi^3_{}}\frac{\textrm{Im}\left[\sum_{\alpha,\beta=e,\mu}^{}\left(U^{\ast}_{\alpha j} U^{\ast}_{\beta j } U^{}_{\beta i} U^{}_{\alpha i} \right)\right]}{\left|U^{}_{ei}\right|^2_{}+\left|U^{}_{\mu i}\right|^2_{}}\nonumber\\
[2mm]
&&
\times \left[S\left(m^{2}_{\nu'^{}_{i}},m^{2}_{\nu'^{}_{j}},M^{2}_{W^{}_R}\right)\right.\nonumber\\
[2mm]
&&\left.+\sum_{a=1,2,3}^{}V^{}_{a}\left(m^{2}_{\nu'^{}_{i}},m^{2}_{\nu'^{}_{j}},M^{2}_{W^{}_R}\right)\right]\,.
\end{eqnarray}
Here the functions $S(x,y,z)$, $V^{}_{1}(x,y,z)$,  $V^{}_{2}(x,y,z)$ and $V^{}_{3}(x,y,z)$ can be respectively deduced from the self-energy correction, the first, second and third vertex correction in Fig. \ref{mndecay}. For example, we find 
\begin{eqnarray}
\label{svcp}
S&=&V_1^{}=3V_2^{}=3V_3^{}= \frac{m^{}_{\nu'^{}_{i}}}{m^{}_{\nu'^{}_{j}}}\frac{m^{4}_{\nu'^{}_{i}}}{m^{4}_{\nu'^{}_{x}}}r^2_{} =\frac{m^{}_{\nu^{}_{i}}}{m^{}_{\nu^{}_{j}}}\frac{m^{4}_{\nu^{}_{i}}}{m^{4}_{\nu^{}_{x}}}r^2_{}\nonumber\\
[2mm]
&&\textrm{for}~~m^{2}_{\nu'^{}_{i}}\ll m^{2}_{\nu'^{}_{j}}\,,~ M^{2}_{W^{}_R}\,,\nonumber\\
[2mm]
S&=&\frac{m^{}_{\nu'^{}_{i}}m^{}_{\nu'^{}_{j}}}{m^{2}_{\nu'^{}_{j}}-m^{2}_{\nu'^{}_{i}}}\frac{m^{4}_{\nu'^{}_{i}}}{m^{4}_{\nu'^{}_{x}}}r^2_{}
=\frac{m^{}_{\nu^{}_{i}}m^{}_{\nu^{}_{j}}}{m^{2}_{\nu^{}_{j}}-m^{2}_{\nu^{}_{i}}}\frac{m^{4}_{\nu^{}_{i}}}{m^{4}_{\nu^{}_{x}}}r^2_{}\gg V^{}_{1,2,3}\nonumber\\
[2mm]
&&\textrm{for}~~m^{2}_{\nu'^{}_{i}}\simeq m^{2}_{\nu'^{}_{j}}\ll  M^{2}_{W^{}_R}\,.
\end{eqnarray}

When the mirror neutrinos $\nu'^{}_{i}$ go out of equilibrium at a temperature $T_D^{}$, their decays can generate a lepton asymmetry stored in the mirror charged leptons $e'$ and $\mu'$, i.e.
\begin{eqnarray}
\label{lasymmetry}
\frac{n^{}_L}{s} &\simeq&  \sum^{}_{i}\varepsilon^{}_{\nu'^{}_{i}}\left(\frac{n^{\textrm{eq}}_{\nu'^{}_{i}}}{s}\right)\left|_{T_D^{}}^{}\right.\nonumber\\
[2mm]
&&\textrm{for~quasi-degenerate~neutrino~spectrum},\nonumber\\
[2mm]
\frac{n^{}_L}{s} &\simeq&  \varepsilon^{}_{\nu'^{}_{x}}\left( \frac{n^{\textrm{eq}}_{\nu'^{}_{x}}}{s}\right)\left|_{T_D^{}}^{}\right.\nonumber\\
&&\textrm{for~hierarchical~neutrino~spectrum}.
\end{eqnarray}
Here $n^{\textrm{eq}}_{\nu'^{}_{i}}=2\left[m^{}_{\nu'^{}_{i}}T/(2\pi)\right]^{3/2}_{}\exp{\left(-m^{}_{\nu'^{}_{i}}/T\right)}$ and $s=2\pi^2_{}g^{}_\ast T^3_{}/45$ respectively are the equilibrium number density and the entropy density with $g_\ast^{}=\mathcal{O}(100)$ being the relativistic degrees of freedom. Subsequently, the mirror charged leptons $e'$ and $\mu'$ can efficiently decay into the ordinary charged leptons $e,\mu,\tau$ and the scalar $\chi$ before the $SU(2)^{}_L$ sphaleron processes stop working, 
\begin{eqnarray}
e'^{\pm}_{},~\mu'^{\pm}_{}\longrightarrow \sum_{\alpha=e,\mu,\tau}^{}e^{\pm}_{R\alpha} + \chi\,.
\end{eqnarray}
The mirror lepton asymmetry (\ref{lasymmetry}) thus can be partially converted to a baryon asymmetry stored in the ordinary quarks, i.e.     
\begin{eqnarray}
\label{basymmetry}
\frac{n^{}_B}{s} =-\frac{28}{79}\frac{n^{}_L}{s} \,.
\end{eqnarray}
This means the final baryon asymmetry can be well described by the ordinary neutrino mass matrix $m^{}_{\nu}$, up to an overall factor.

We now estimate the temperature $T_D^{}$. For this purpose, we need consider the annihilations of the mirror neutrinos into the mirror charged fermions, $\nu'^{}_{i}+\nu'^c_{i}\rightarrow e'+e'^c_{}$, $\mu'+\mu'^c_{}$, $u'+u'^c_{}$ and $d'+d'^c_{}$. The annihilation cross section is \cite{kt1990}
\begin{eqnarray}
\label{anncs}
\langle\sigma_{A}^{}v^{}_{\textrm{rel}}\rangle&=& \frac{M^{4}_{W^{}_L}}{M^{4}_{W^{}_R}}\frac{G^{2}_{F}m^{2}_{\nu'^{}_{i}}}{2\pi} \left\{2\left[\left(-1+4\tan^2_{}\theta^{}_W\right)^2_{}+1\right]\right.\nonumber\\
[2mm]
&&+3\left[\left(-1+\frac{4}{3}\tan^2_{}\theta^{}_W\right)^2_{}+1\right]\nonumber\\
[2mm]
&&\left.+3\left[\left(1-\frac{8}{3}\tan^2_{}\theta^{}_W\right)^2_{}+1\right]\right\} \frac{8}{3}\langle v^2_{\textrm{rel}}\rangle\nonumber\\
[2mm]
& \simeq & \frac{64}{\pi}\left(2-5\tan^2_{}\theta^{}_W+ \frac{22}{3}\tan^4_{}\theta^{}_W\right) \frac{M^{4}_{W^{}_L}}{M^{4}_{W^{}_R}}\nonumber\\
[2mm]
&&\times G^{2}_{F}m^{}_{\nu'^{}_{i}} T~~\textrm{for}~~m^{}_{\nu'^{}_{i}} \ll M^{}_{Z^{}_R}\,,~M^{}_{W^{\pm}_R}\,.
\end{eqnarray}
The interaction rate then should be smaller than the Hubble constant,
\begin{eqnarray}
\label{annrate}
\left[\Gamma=n^{\textrm{eq}}_{\nu'^{}_{i}}\langle\sigma_{A}^{}v^{}_{\textrm{rel}}\rangle<H(T)=\left(\frac{8\pi^{3}_{}g_{\ast}^{}}{90}\right)^{\frac{1}{2}}_{}
\frac{T^{2}_{}}{M_{\textrm{Pl}}^{}}\right]\left|_{T_D^{}}^{}\right..
\end{eqnarray}
Here $M_{\textrm{Pl}}^{}\simeq 1.22\times 10^{19}_{}\,\textrm{GeV}$ is the Planck mass. As an example, we can obtain 
\begin{eqnarray}
\label{par1}
&&T_D^{}\simeq 0.2\,m^{}_{\nu'^{}_{i}}\,,~~\left( \frac{n^{\textrm{eq}}_{\nu'^{}_{i}}}{s}\right)\left|_{T_D^{}}^{}\right.\simeq 2\times 10^{-4}_{}\nonumber\\
[2mm]
&&\quad \quad \textrm{for}~~M^{}_{W^{}_R}=3.3\,m_{\nu'^{}_{i}}^{}=10^{12}_{}\,\textrm{GeV}\,.
\end{eqnarray}
The result is similar for the scattering processes, $\nu'^{}_{i} + e'^{+}_{} (\mu'^{+}_{})\rightarrow u'^{}+d'^{c}_{}$, $\nu'^{}_{i} + d'^{}_{} \rightarrow e'^{-}_{} (\mu'^{-}_{})+u'$, $\nu'^{}_{i} + u'^{c}_{} \rightarrow e'^{-}_{} (\mu'^{-}_{})+d'^{c}_{}$. For the parameter choice (\ref{par1}), the mirror neutrino decays has been out of equilibrium below the temperature $T\simeq m_{\nu'^{}_{i}}^{}$. Actually, we have 
\begin{eqnarray}
\!\!\!\!\!\!\!\!&&\frac{\Gamma^{}_{\nu'^{}_{i}}}{H(T)}\left|_{T=m_{\nu'^{}_{i}}^{}}^{}\right.\lesssim 0.08~~\textrm{for}~~M^{}_{W^{}_R}=3.3\,m_{\nu'^{}_{i}}^{}=10^{12}_{}\,\textrm{GeV}\,.\nonumber\\
\!\!\!\!\!\!\!\!&&
\end{eqnarray}
We further consider the quasi-degenerate neutrino spectrum $m^{}_{\nu'^{}_{1}}\lesssim m^{}_{\nu'^{}_{2}}\lesssim m^{}_{\nu'^{}_{3}} \simeq 0.2 \,\textrm{eV}$ and fix the Majorana CP phases $\alpha_{1,2}^{}$ to be $\alpha_{1}^{}=\alpha_{2}^{}=\pi/2$. In this case, the mirror lepton asymmetry from the $\nu'^{}_{1,2}$ decays should dominate over that from the $\nu'^{}_{3}$ decays because the resonant enhancement exists in the CP asymmetries $\varepsilon^{}_{\nu'^{}_{1,2}}$ other than the CP asymmetry $\varepsilon^{}_{\nu'^{}_{3}}$, i.e.
\begin{eqnarray}
\varepsilon^{}_{\nu'^{}_{1}}&\simeq& \frac{g^4_{}}{2^{9}_{}\pi^3_{}}\frac{s_{12}^{}c_{12}^{}s_{23}^{}c_{23}^{}c_{13}^2 s_{13}^{}\sin\delta }{c_{12}^2 c_{13}^2 + s_{12}^2 c_{23}^2}\frac{m_{\nu^{}_1}^{}m_{\nu^{}_2}^{}}{m_{\nu^{}_2}^2- m_{\nu^{}_1}^2}r^2_{}\,,\nonumber\\
[2mm]
\varepsilon^{}_{\nu'^{}_{2}}&\simeq& \frac{g^4_{}}{2^{9}_{}\pi^3_{}}\frac{s_{12}^{}c_{12}^{}s_{23}^{}c_{23}^{}c_{13}^2 s_{13}^{}\sin\delta }{s_{12}^2 c_{13}^2 + c_{12}^2 c_{23}^2}\frac{m_{\nu^{}_1}^{}m_{\nu^{}_2}^{}}{m_{\nu^{}_2}^2- m_{\nu^{}_1}^2}r^2_{}\,,\nonumber\\
[2mm]
\varepsilon^{}_{\nu'^{}_{3}}&\ll& \varepsilon^{}_{\nu'^{}_{1,2}}~~\textrm{with}~~r=0.1 \,.
\end{eqnarray}
The final baryon asymmetry then should be
\begin{eqnarray}
\frac{n^{}_B}{s} &\simeq& -\frac{28}{79}\left[\varepsilon^{}_{\nu'^{}_{1}}\left(\frac{n^{\textrm{eq}}_{\nu'^{}_{1}}}{s}\right)\left|_{T_D^{}}^{}\right. +\varepsilon^{}_{\nu'^{}_{2}}\left(\frac{n^{\textrm{eq}}_{\nu'^{}_{2}}}{s}\right)\left|_{T_D^{}}^{}\right.\right]\nonumber\\
[2mm]
&\simeq&10^{-10}_{}\left(\frac{\sin\delta}{-0.27}\right) \,,\end{eqnarray}
which is able to account for the observations.

Note the lepton-number-violating interactions for generating the Majorana masses of the ordinary and mirror neutrinos should go out of equilibrium before the above leptogenesis epoch. Fortunately the related processes can decouple at a very high temperature \cite{fy1990}, 
\begin{eqnarray}
T_F^{}&=&10^{12}_{}\,\textrm{GeV}\left[\frac{0.04\,\textrm{eV}^2_{}}{\sum_{i}^{}m^{2}_{\nu^{}_i}}\right]\left[\frac{v^{}_{2}}{v}\right]^4_{}\nonumber\\
[2mm]
&&\textrm{for}~~T_F^{}\ll M_{N}^{}\,.
\end{eqnarray}
The mirror neutrinos should begin to decay below this temperature $T_F^{}$. The parameter choice (\ref{par1}) can fulfil this requirement. 

\section{Strong CP problem and dark matter}

The present model give a non-perturbative QCD Lagrangian as follows,
\begin{eqnarray}
\!\!\!\!&&\!\!\!\!\mathcal{L}_{QCD}^{}\supset-\bar{\theta}\frac{g^2_3}{32\pi^2_{}}G\tilde{G}~~\textrm{with}~~
\bar{\theta}=\theta-\textrm{Arg}\textrm{Det} (M_u^{} M_d^{})\,,\nonumber\\
\!\!\!\!&&\!\!\!\!
\end{eqnarray}
where $\theta$ is from the QCD $\Theta$-vacuum while $M_u^{}$ and $M_d^{}$
are the mass matrices of the down-type and up-type quarks,
\begin{eqnarray}
M_d^{}&=&\left[\begin{array}{cc}\frac{1}{\sqrt{2}}\bar{y}_d^{}v^{}_1&0\\
[2mm]
0&\frac{1}{\sqrt{2}}\bar{y}_d^{\dagger}v'^{}_1\end{array}\right],\nonumber\\
[2mm]
M_u^{}&=&
\left[\begin{array}{cc}\frac{1}{\sqrt{2}}\bar{y}_u^{}v^{}_2&0\\
[2mm]
0&\frac{1}{\sqrt{2}}\bar{y}_u^{\dagger}v'^{}_{2}\end{array}\right].
\end{eqnarray}
When the $\theta$-term is removed as a result of the
parity invariance, the real determinants
$\textrm{Det}(M_d^{})$ and $\textrm{Det}(M_u^{})$ will lead to a
zero $\textrm{Arg}\textrm{Det} (M_u^{} M_d^{})$. We hence can obtain
a vanishing strong CP phase $\bar{\theta}$ at tree level \cite{bm1989}.

The model also contains a stable scalar $\chi$ because of the unbroken $Z^{}_2\times Z^{}_2$ symmetry. This scalar can annihilate into the ordinary species through the Higgs portal interaction. This simple dark matter scenario has been studied in a lot of literatures \cite{sz1985,ht2016,arcadi2017,athron2017}.

\section{Summary}

In this paper we have shown the spontaneous left-right symmetry breaking can automatically break the parity symmetry motivated by solving the strong CP problem. As a result, the mass matrices of the mirror fermions and their ordinary partners can have a same structure although their scales are allowed very different. Through the $SU(2)_R^{}$ gauge interactions, a mirror Majorana neutrino can decay into a mirror charged lepton and two mirror quarks. Consequently we can obtain a lepton asymmetry stored in the mirror charged leptons. The Yukawa couplings of the mirror and ordinary charged fermions to the dark matter scalar then can transfer the mirror lepton asymmetry to an ordinary lepton asymmetry. The $SU(2)^{}_L$ sphaleron processes eventually can realize the conversion of the lepton asymmetry to the baryon asymmetry. In this novel leptogenesis scenario, the cosmic baryon asymmetry can be well described by the Majorana neutrino mass matrix up to an overall factor. 

\textbf{Acknowledgement}:  This work was supported by the National Natural Science Foundation of China under Grant No. 11675100, the Recruitment Program for Young Professionals under Grant No. 15Z127060004, the Shanghai Jiao Tong University under Grant No. WF220407201, the Shanghai Laboratory for Particle Physics and Cosmology, and the Key Laboratory for Particle Physics, Astrophysics and Cosmology, Ministry of Education.

\appendix

\section{Useful arithmetic of PMNS elements}

\begin{eqnarray}
&&\textrm{Im}\left(U_{\mu 2}^{\ast} U_{e 2}^{\ast} U_{\mu 1}^{} U_{e 1}^{} \right)= -\textrm{Im}\left(U_{\mu 1}^{\ast} U_{e 1}^{\ast} U_{\mu 2}^{} U_{e 2}^{} \right)\nonumber\\
[2mm]
&=&s_{12}^2 c_{12}^2 c_{13}^2  (s_{23}^2 s _{13}^2 - c_{23}^2 )\sin(\alpha_1^{}-\alpha_2^{}) \nonumber\\
[2mm]
&&+ s_{12}^{}c_{12}^{}s_{23}^{}c_{23}^{}c_{13}^2 s_{13}^{} [s_{12}^2 \sin(\alpha_1^{}+\alpha_2^{}-\delta)\nonumber\\
[2mm]
&&-c_{12}^2 \sin(\alpha_1^{}+\alpha_2^{}+\delta)]\,,\nonumber\\
[5mm]
&&\textrm{Im}\left(U_{\mu 3}^{\ast} U_{e 3}^{\ast} U_{\mu 1}^{} U_{e 1}^{} \right)= -\textrm{Im}\left(U_{\mu 1}^{\ast} U_{e 1}^{\ast} U_{\mu 3}^{} U_{e 3}^{} \right)\nonumber\\
[2mm]
&=&-s_{12}^{}c_{12}^{}s_{23}^{}c_{23}^{}c_{13}^2 s_{13}^{}  \sin(\alpha_1^{}+\delta)\nonumber\\
[2mm]
&&-c_{12}^2s_{23}^2c_{13}^2 s_{13}^2 \sin(\alpha_1^{}+2\delta)\,,\nonumber\\
[5mm]
&&\textrm{Im}\left(U_{\mu 3}^{\ast} U_{e 3}^{\ast} U_{\mu 2}^{} U_{e 2}^{} \right)= -\textrm{Im}\left(U_{\mu 2}^{\ast} U_{e 2}^{\ast} U_{\mu 3}^{} U_{e 3}^{} \right)\nonumber\\
[2mm]
&=&s_{12}^{}c_{12}^{}s_{23}^{}c_{23}^{}c_{13}^2 s_{13}^{}  \sin(\alpha_2^{}+\delta)\nonumber\\
[2mm]
&&-s_{12}^2 s_{23}^2 c_{13}^2 s_{13}^2 \sin(\alpha_2^{}+2\delta)\,.
\end{eqnarray}

\begin{eqnarray}
&&|U^{}_{e1}|^2_{}+|U^{}_{\mu 1}|^2_{}\nonumber\\
[2mm]
&=&c_{12}^2 c_{13}^2 + [s_{12}^{}c_{23}^{}\cos(\alpha_1^{}/2)+c_{12}^{}s_{23}^{}s_{13}^{}\cos(\alpha_1^{}/2+\delta)]^2_{}\nonumber\\
[2mm]
&&+ [s_{12}^{}c_{23}^{}\sin(\alpha_1^{}/2)+c_{12}^{}s_{23}^{}s_{13}^{}\sin(\alpha_1^{}/2+\delta)]^2_{}\,,\nonumber\\
[5mm]
&&|U^{}_{e2}|^2_{}+|U^{}_{\mu 2}|^2_{}\nonumber\\
[2mm]
&=& s_{12}^2 c_{13}^2 + [c_{12}^{}c_{23}^{}\cos(\alpha_2^{}/2)-s_{12}^{}s_{23}^{}s_{13}^{}\cos(\alpha_2^{}/2+\delta)]^2_{}\nonumber\\
[2mm]
&&+ [c_{12}^{}c_{23}^{}\sin(\alpha_2^{}/2)-s_{12}^{}s_{23}^{}s_{13}^{}\sin(\alpha_2^{}/2+\delta)]^2_{}\,,\nonumber\\
[5mm]
&&|U^{}_{e3}|^2_{}+|U^{}_{\mu 3}|^2_{}=s_{13}^2 + s_{23}^2 c_{13}^2\,.
\end{eqnarray}

\end{document}